\documentclass[10pt, conference, compsocconf]{IEEEtran}
\usepackage{amsmath,amssymb,amsthm,graphicx,epstopdf, cite}
\usepackage[tight,footnotesize]{subfigure}
\usepackage{color}

\def \feasiblesize {\mathcal X}
\def \controlproblem {\mathcal C}
\def \heuristiccontrol {\mathcal{H}_{\text{s}}}
\def \heuristicsizing {\mathcal{H}_{\text{osz}}}
\def \heuristiconline {\mathcal{H}_{\text{rt}}}

\newcommand{\bq}{\begin{IEEEeqnarray}{rCl}}
\newcommand{\eq}{\end{IEEEeqnarray}}

\theoremstyle{remark}
\newtheorem*{heu}{Heuristic}

\ifCLASSINFOpdf
\else
\fi
\begin{document}
%
\title{Optimal Sizing of Voltage Control Devices for Distribution Circuit with Intermittent Load}

\author{\IEEEauthorblockN{Changhong Zhao}
\IEEEauthorblockA{Department of Electrical Engineering\\
California Institute of Technology\\
Pasadena, California 91125, USA\\
Email: czhao@caltech.edu}
\and
\IEEEauthorblockN{Michael Chertkov}
\IEEEauthorblockA{Theoretical Division\\
Los Alamos National Laboratory\\
Los Alamos, NM 87545, USA\\
Email: chertkov@lanl.gov}
\and
\IEEEauthorblockN{Scott Backhaus}
\IEEEauthorblockA{Materials Physics \& Applications Division\\
Los Alamos National Laboratory\\
Los Alamos, NM 87545, USA\\
Email: backhaus@lanl.gov}}


%


\maketitle

\begin{abstract}

We consider joint control of a switchable capacitor and a D-STATCOM for voltage regulation in a distribution circuit with intermittent load. The control problem is formulated as a two-timescale optimal power flow problem with chance constraints, which minimizes power loss while limiting the probability of voltage violations due to fast changes in load.
The control problem forms the basis of an optimization problem which determines the sizes of the control devices by minimizing sum of the expected power loss cost and the capital cost. We develop computationally efficient heuristics to solve the optimal sizing problem and implement real-time control. Numerical experiments on a circuit with high-performance computing (HPC) load show that the proposed sizing and control schemes significantly improve the reliability of voltage regulation on the expense of only a moderate increase in cost.

\end{abstract}


%
\IEEEpeerreviewmaketitle

\section{Introduction}
The effect of intermittent generation or load on the quality of voltage regulation in distribution circuits has recently received significant attention \cite{Turitsyn2011options}. Much of this focus has been on the design of control algorithms for modifying the reactive power injections along a distribution circuit to maintain voltage within acceptable bounds. The reactive power injections may be derived from spatially concentrated sources such as fixed and switchable capacitors \cite{baran1989optimalsizing, baran1989optimalplacement} and D-STATCOMs \cite{moreno2007power}, or distributed sources such as photovoltaic (PV) inverters \cite{liu2008distribution, smith2011smart, turitsyn2010local, Turitsyn2011options, vsulc2013optimal, kundu2013distributed, farivar2011inverter, farivar2012optimal} and other distributed generation inverters \cite{joos2000potential}.
There are also various mechanisms to jointly control two or more kinds of reactive power sources, e.g., \cite{grainger1985volt1, civanlar1985volt2, civanlar1985volt3, baldick1990efficient} for switchable capacitors and tap-changing voltage regulators, \cite{farivar2011inverter} for switchable capacitors and inverters, and \cite{senjyu2008optimal} for capacitors, reactors and static var compensators, et cetera.
Meanwhile the problem of optimal placement and sizing of capacitors has been extensively studied using analytical methods \cite{grainger1981optimum, salama1985control, grainger1985volt1, civanlar1985volt2, civanlar1985volt3}, numerical programming \cite{dura1968optimum, baran1989optimalsizing, baran1989optimalplacement}, and probabilistic meta-heuristics like simulated annealing \cite{ananthapadmanabha1996knowledge} and genetic algorithm \cite{boone1993optimal, sundhararajan1994optimal}; see \cite{ng2000classification} for more.

However, most of the work above considered voltage control either at a slow timescale (e.g., using switchable capacitors and tap-changing regulators) or at a fast timescale (e.g., using inverters), without combining the controls at two timescales for better performance. An exception is \cite{farivar2011inverter}, in which a two-timescale control problem was formulated with switchable capacitors at slow timescale and inverters at fast timescale. However the assumption in \cite{farivar2011inverter} that the aggregate load changes gradually over time and it is thus well predicted does not hold for the highly intermittent load we consider in this paper.
Moreover, absent in much of the work above are methods to size different sources of reactive power when they are jointly controlled.
To the best of our knowledge, our work in this paper is the first to optimally control and size multiple kinds of reactive power sources working at different timescales by incorporating statistical characterization of rapid and large load changes over time.

\begin{figure}[!t]
\centering
\includegraphics[height=5.4 cm]{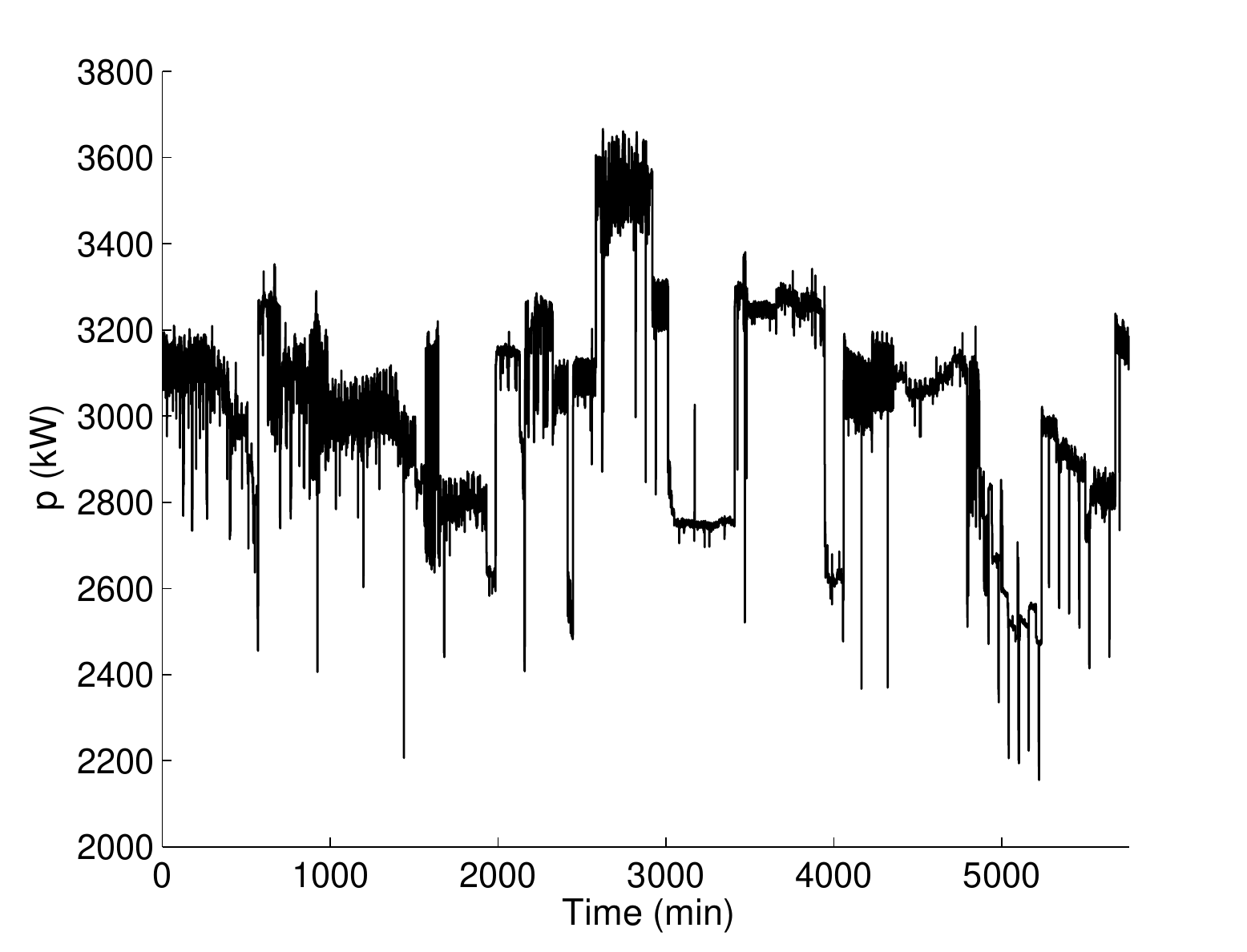}
\caption{A four-day segment of the measured real power consumption time-series for the HPC load used in this paper. The real power was sampled every five seconds.}\label{fig:p_trace}
\end{figure}

While small distributed PV generation has stimulated much of the research in this area, large and highly intermittent loads or generation can create similar, and perhaps more difficult, problems.  One such example is a large (several MW) PV generator.  However, the motivating example for this work is a high-performance computing (HPC) load. Power consumption of a modern HPC load can easily swing several MW in a few seconds or less.  Fig.~\ref{fig:p_trace} shows a typical time-series trace of the real power consumption of the HPC load that motivated this work.
This set of data show a pattern that large transitions are typically separated by minutes or hours, while relatively small changes continuously occur during the period of time, or \emph{stage}, between two consecutive large transitions. A voltage control scheme that incorporates smaller, frequently controlled devices and larger, infrequently controlled devices is suitable for such a load pattern. Moreover, conditioned on the current-stage average power, the probability distribution of the next-stage average power reveals information about the direction and size of voltage change that may occur in the next stage. We leverage this information to develop an improved voltage control scheme for distribution circuits with large, rapidly changing loads or generation. We then embed the proposed control into an optimal sizing problem for reactive power sources, which balances the capital cost of the devices with the expected cost due to power losses.

Specifically, we formulate a two-timescale optimization problem for the control of two devices: 1) a relatively inexpensive (potentially large) switchable capacitor that operates primarily at the infrequent transitions between stages, but with some time delay that makes it incapable of quickly correcting large voltage deviations and 2) a far more expensive (potentially smaller) D-STATCOM that operates continuously to follow frequent changes in load and nearly instantaneously after sudden changes in load.
The slow-timescale switchable capacitor control problem is a chance-constrained optimal power flow (OPF) problem which minimizes power loss, regulates the current-stage voltage and, through chance constraints, limits the probability of voltage violations in the next stage caused by a large, sudden change in load. The fast-timescale D-STATCOM control problem is a deterministic-constrained OPF problem that is solved at every time when the load power is sampled.

The control problem above forms the basis of an optimal sizing problem, whose objective is to determine the sizes of the two control devices and a fixed capacitor that minimize the sum of the cost of expected power loss (obtained from results of underlying control problems) and the capital cost. Through simplifications of the underlying chance-constrained OPF problems, we develop a computationally efficient heuristic based on simulated annealing to solve the sizing problem. The process of developing this heuristic also helps us develop another heuristic for real-time implementation of the proposed voltage control.

With a realistic HPC load, we solve the sizing problem using the proposed heuristic and run simulations of real-time control to demonstrate that the proposed control and sizing schemes achieve the desired tradeoff between the reliability of voltage regulation and cost efficiency. We also perform a parametric study to investigate the impact of changing the chance constraints on the performance of the proposed control and on the total (capital plus power-loss-induced) cost. We note again that the methods in this paper are applicable to other scenarios, e.g., intermittent renewable generation, where the statistics of fast load or generation changes can be characterized, for control and system planning.

The rest of this paper is organized as follows.  Section~\ref{sec:model} describes the model of the distribution circuit and the HPC load.  Section~\ref{sec:formulation} formulates the control and sizing problems. Section~\ref{sec:allheuristics} describes our heuristics to solve the sizing problem and implement real-time control.  Section~\ref{sec:results} presents the results of optimal sizing as well as simulations of the real-time control. Finally, Section~\ref{sec:conclusions} discusses our conclusions and directions for future work.

\section{Modeling of HPC distribution circuit}\label{sec:model}

\subsection{Circuit model}\label{subsec:circuit}

\begin{figure}[!t]
\centering
\includegraphics[height=4.5 cm]{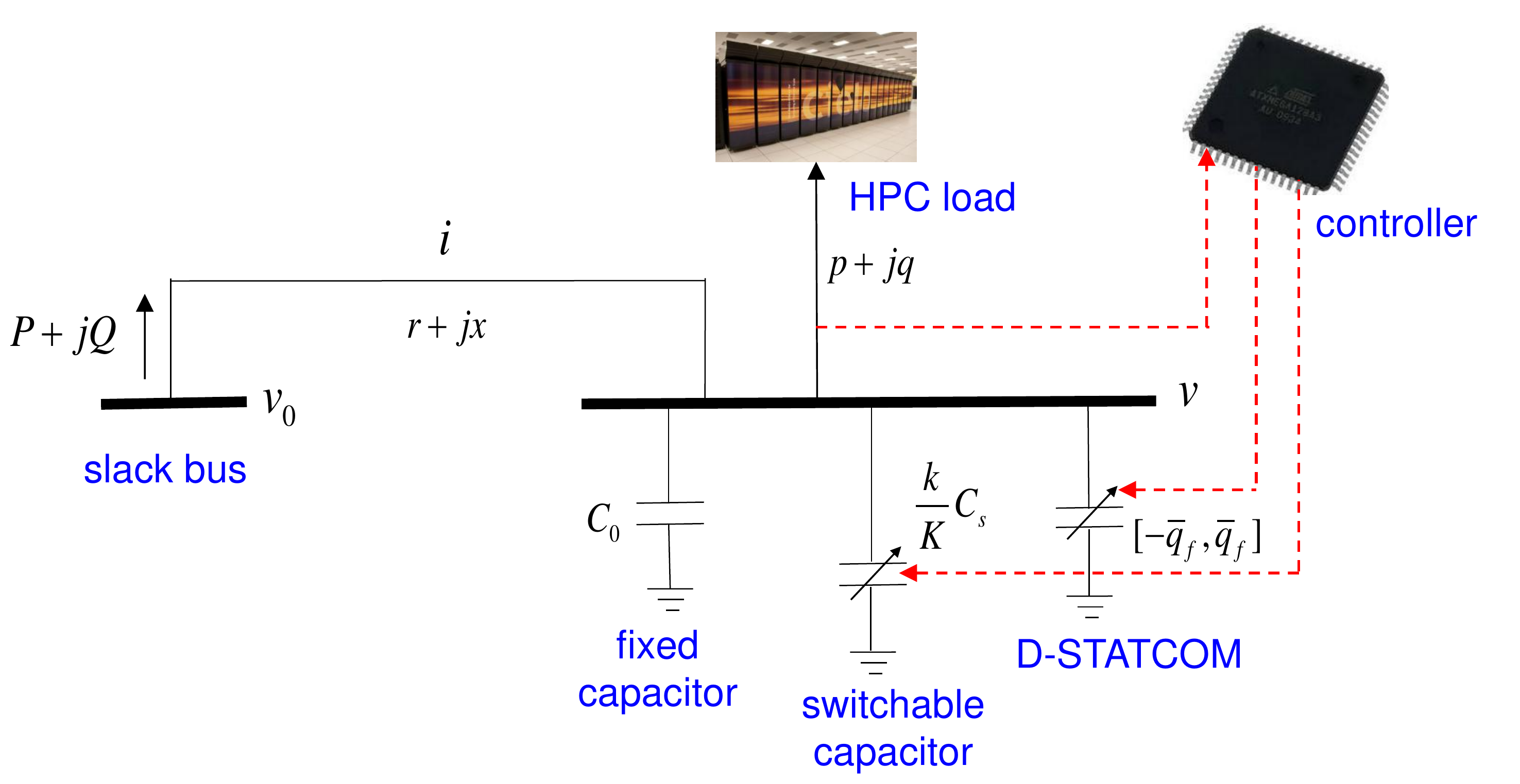}
\caption{Schematic of the simplified circuit with resistance $r$ and reactance $x$ supplying an HPC load at a voltage magnitude of $v$ from a slack bus at a voltage magnitude of $v_0$. Real power $P$ and reactive power $Q$ flow from the slack bus with real power $p$ and reactive power $q$ consumed by the load. The current magnitude in the circuit is $i$. Three reactive power sources and a controller are installed near the load, as explained in Section \ref{subsec:circuit}. Black lines are actual circuit lines and red lines represent signal flows.
}
\label{fig:HPC_config}
\end{figure}

Modern HPC platforms are typically supplied by multiple distribution circuits to ensure a redundant power supply.  Here we simplify this configuration by considering a single radial circuit (see Fig.~\ref{fig:HPC_config}) with a single HPC load concentrated at its end.
A single branch of resistance $r$ and reactance $x$ connects the HPC load to a slack bus of fixed voltage magnitude and phase angle. Without loss of generality, the voltage phase angle at the slack bus is set to zero. We denote the voltage magnitude at the slack bus by $v_0$, the voltage magnitude at the HPC load by $v$, the current magnitude on the branch by $i$, the real and reactive powers sent from the slack bus by $P$ and $Q$, and the real and reactive power consumptions of the HPC load by $p$ and $q$. We assume
\bq
q = \phi p \label{eq:DF_phi}
\eq
where $\phi$ is a positive constant. The constant load power factor represented by \eqref{eq:DF_phi} is typical of the HPC power consumptions.

The following devices are installed at the end of the circuit for voltage regulation. First, a fixed capacitor with capacitance $C_0$ injects voltage-dependent reactive power $v^2 f_0  C_0$, where $f_0$ is the frequency of the circuit and is assumed to be constant. Second, a switchable capacitor can switch to a small number $K+1$ of discrete values of capacitance $c_s \in \left\{\frac{k}{K}C_s| k = 0,...,K\right\}$. In most distribution circuits, capacitors switch only a few times each day to adapt to the gradual changes of the aggregate load, due to their limited life cycle \cite{farivar2011inverter}. In the specific system we consider, however, the capacitor switches more frequently than a few times a day to adapt to the highly intermittent HPC load, but still much less frequently than the changes in HPC load to avoid excessive wear and tear. Moreover, the capacitor cannot switch as fast as the changes in HPC load since the mechanical switching time of the capacitor will delay the implementation of control by many AC cycles even if a change in load is detected instantaneously. In this work, as explained below, we determine when to switch the capacitor in real time based on the actual load and take into account the switching time delay. Third, a D-STATCOM injects reactive power $q_f$ ranging continuously within a preset range $\left[-\overline q_f,~\overline q_f\right]$. A D-STATCOM is much more expensive than a capacitor with the same maximum reactive power injection, but can respond within an AC cycle to a change in load and does not suffer from wear and tear from frequent changes in $q_f$. We call $\left(C_0, C_s, \overline q_f\right)$ the \emph{sizes} of reactive power sources and $\left(c_s,q_f\right)$ the control variables. A real-time feedback controller is installed at the load side of the circuit, which measures variables like load power, voltage and current, takes them as input, and computes the values of control variables and sends them to various control devices.

Suppose the parameter values $\left(r,x,f_0, v_0, \phi\right)$ are given and fixed. Then, incorporating \eqref{eq:DF_phi}, the real power $p$ of HPC load, the size $C_0$ of fixed capacitor, the control $\left(c_s, q_f\right)$ and the state variables $\left(v,i,P,Q\right)$ satisfy
\bq
i^2 &=&\frac{P^2+Q^2}{v_0^2} \label{eq:DF_l}\\
P&=&p+i^2r, \label{eq:DF_P}\\
Q&=&\phi p- v^2 f_0  \left(C_0 + c_s \right) - q_f+i^2x,  \label{eq:DF_Q}\\
v^2&=&v_0^2 -2(r P+x Q)+ i^2 (r^2+x^2). \label{eq:DF_v}
\eq
The equations are known as the DistFlow equations \cite{baran1989optimalsizing}. Note that $i^2r$ and $i^2x$ in \eqref{eq:DF_P} and \eqref{eq:DF_Q} are respectively the real and reactive power losses. With $(p,C_0,c_s,q_f)$ specified, the four variables $\left(v,i,P,Q\right)$ can be solved from the four equations \eqref{eq:DF_l}--\eqref{eq:DF_v}. Indeed, there are two solutions (both with nonnegative values of $v$ and $i$), one with $v$ close to $v_0$, small $i$ and hence small power loss, and the other with $v$ close to zero, large $i$ and hence large power loss. We only care about the first one and take it as the unique solution because we desire good voltage regulation and minimized power loss \cite{chiang1990existence}, \cite{andersson2004modelling}. Hence $v$, $i$, $P$, $Q$ can be written as functions of $(p,C_0,c_s,q_f)$, e.g., $v = v\left(p,C_0,c_s,q_f\right)$ and $i = i\left(p,C_0,c_s,q_f\right)$.\footnote{We abuse the notations by using $v$ and $i$ to denote either the variables or the functions, depending on the context.}

\subsection{HPC load model}\label{subsec:loadmodel}

Equations \eqref{eq:DF_l}--\eqref{eq:DF_v} describe the behavior of the circuit at a particular instant. In practice, the real power $p$ of HPC load constantly changes over time, so may the control $(c_s,q_f)$ and state variables $\left(v,i,P,Q\right)$. Here, we focus on characterizing the changes in $p$ over time. As an example, we consider the real power usage recorded at a large HPC platform at Los Alamos National Laboratory.

Fig. \ref{fig:p_trace} shows the time-series trace of $p$ over four days, sampled every 5 seconds. The minimum and maximum values of the trace in the four days are $\underline p = 2150~\text{kW}$ and $\overline p = 3650~\text{kW}$, and we assume $p \in \left[\underline p,~\overline p\right]$ always holds. Let $\tau \in \mathbb{N}_0 =\left\{0,1,...\right\}$ index the time at which the (real) power is sampled and $p(\tau)$ denote the power sampled at time $\tau$. We see from Fig.~\ref{fig:p_trace} that $\left| p(\tau+1) - p(\tau) \right|$ are relatively small (less than 200 kW) for most of the time while large changes from $p(\tau)$ to $p(\tau+1)$ are infrequent and usually separated by minutes or even hours.

To capture this pattern, we divide the sequence $\left\{p(\tau),~\tau\in \mathbb{N}_0\right\}$ into \emph{stages}. A stage, indexed by $t\in \mathbb{N}_0$, is a subsequence $\left\{p(\tau_t), p(\tau_t +1),...,p(\tau_{t+1}-1)\right\}$ where $\tau_t$ and $\tau_{t+1}$ are the times of two consecutive large changes in $p$. The average power of stage $t$ is $p[t]:=\frac{p(\tau_t) + p(\tau_t +1) +...+p(\tau_{t+1}-1)}{T_t}$ where $T_t: = \tau_{t+1} - \tau_{t}$ is the duration of stage $t$. In Section \ref{sec:heuristic_for_online_implementation} we propose a method to determine the durations and average powers of stages for a given sequence $\left\{p(\tau),~\tau\in \mathbb{N}_0\right\}$.

\begin{figure*}[!t]
\centering
\subfigure[]
{\includegraphics[height=4.2 cm]{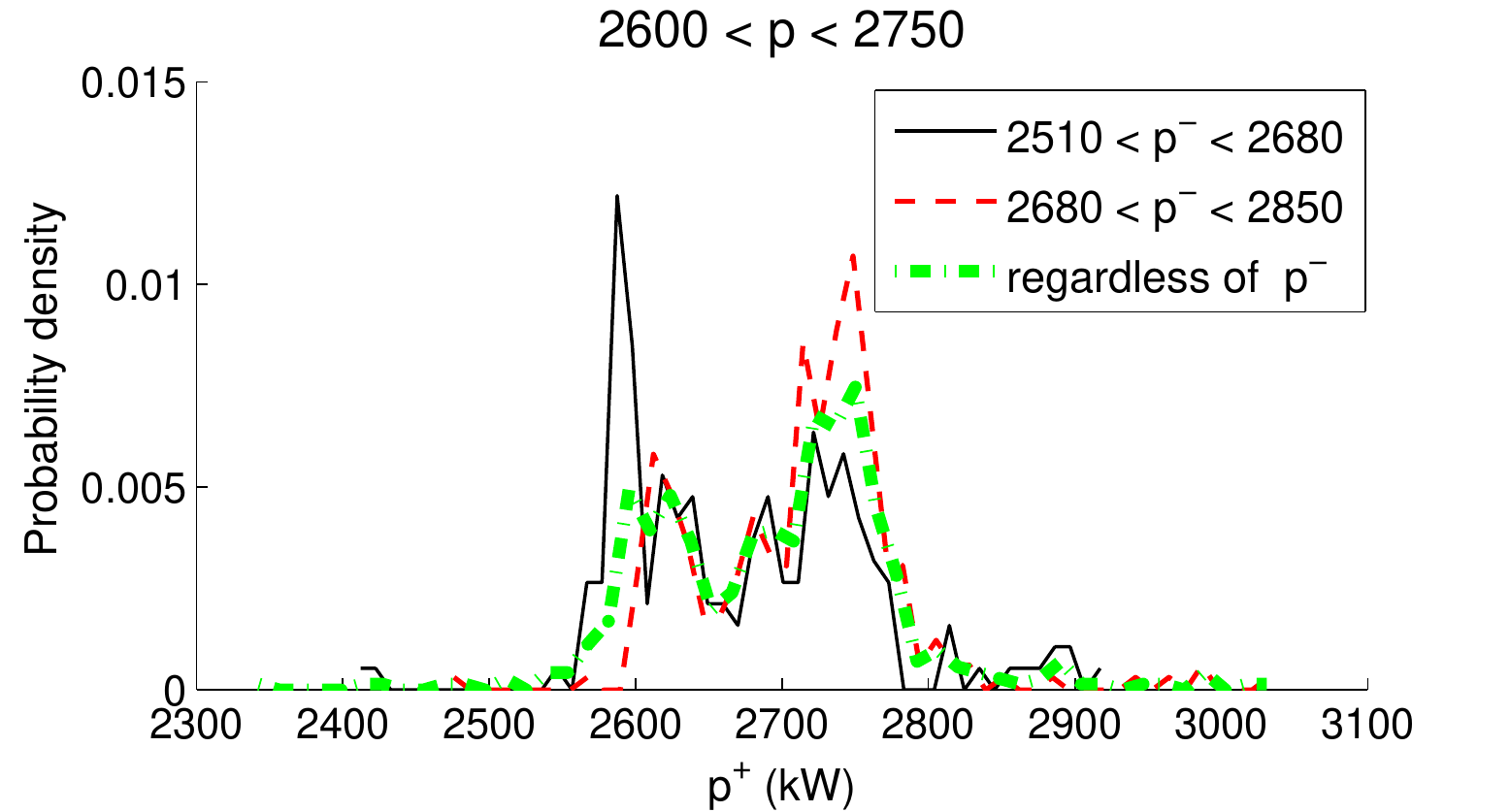}\label{fig:markov_a}}
\hfil
\subfigure[]
{\includegraphics[height=4.2 cm]{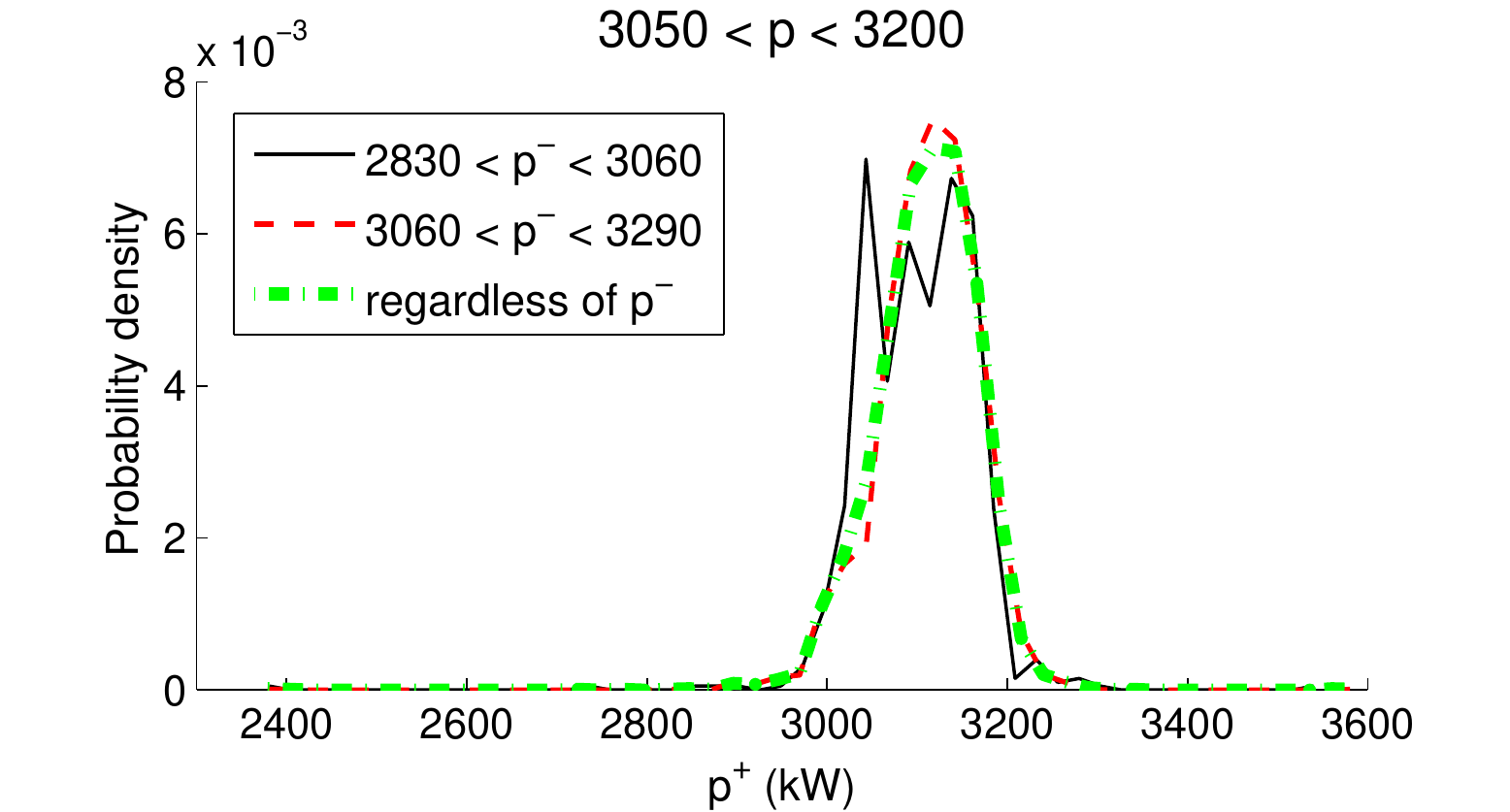}\label{fig:markov_b}}
\caption{Examples of probability density of $p^+$ conditioned on $p$ and $p^-$. Subfigures (a) and (b) are for different $p$, and the legends label different $p^-$. To collect sufficiently many samples to plot the probability density, $p$ and $p^-$ are taken as some ranges as opposed to numbers.}\label{fig:markov}
\end{figure*}

We assume the sequence of (average load powers of) stages $\left\{p[t],~t \in \mathbb{N}_0\right\}$ forms a first-order homogeneous Markov chain characterized by transition probability $\pi(p^+\left|\right. p)$, where $p$ is the average power of the current stage and $p^+$ is the average power of the next stage. The formal validation of this assumption is our future work, and here we give a partial justification. We determine the sequence of stages from the time-series in Fig. \ref{fig:p_trace} using the method in Section \ref{sec:heuristic_for_online_implementation}, and measure the probability density of the next-stage power $p^+$ conditioned on the current-stage power $p$ and the last-stage power $p^-$, across all the transitions of stage powers. Fig. \ref{fig:markov} shows two examples in which, given $p$, the probability density of $p^+$ is approximately independent of $p^-$, which hints that the sequence of stages may have a first-order homogeneous Markov property. We also assume the sequence of samples $\left\{p(\tau),~\tau \in \mathbb{N}_0\right\}$ has a stationary distribution $\rho(\cdot)$. Note that $\rho$ is different from the stationary distribution of the Markov chain of average stage powers.

\section{Problem formulation}\label{sec:formulation}

\subsection{Voltage control problem}\label{subsec:optimal_var_control}

Suppose the number $K+1$ of switchable capacitor levels is fixed, and the sizes $(C_0,C_s,\overline q_f)$ of reactive power sources are given. We design a real-time voltage control which takes $\left\{p(\tau),~\tau \in \mathbb{N}_0\right\}$ as input and computes optimal output $\{(c_s^*(\tau),q_f^*(\tau)),~\tau \in \mathbb{N}_0\}$.

\begin{figure}[!t]
\centering
\includegraphics[height=5.2 cm]{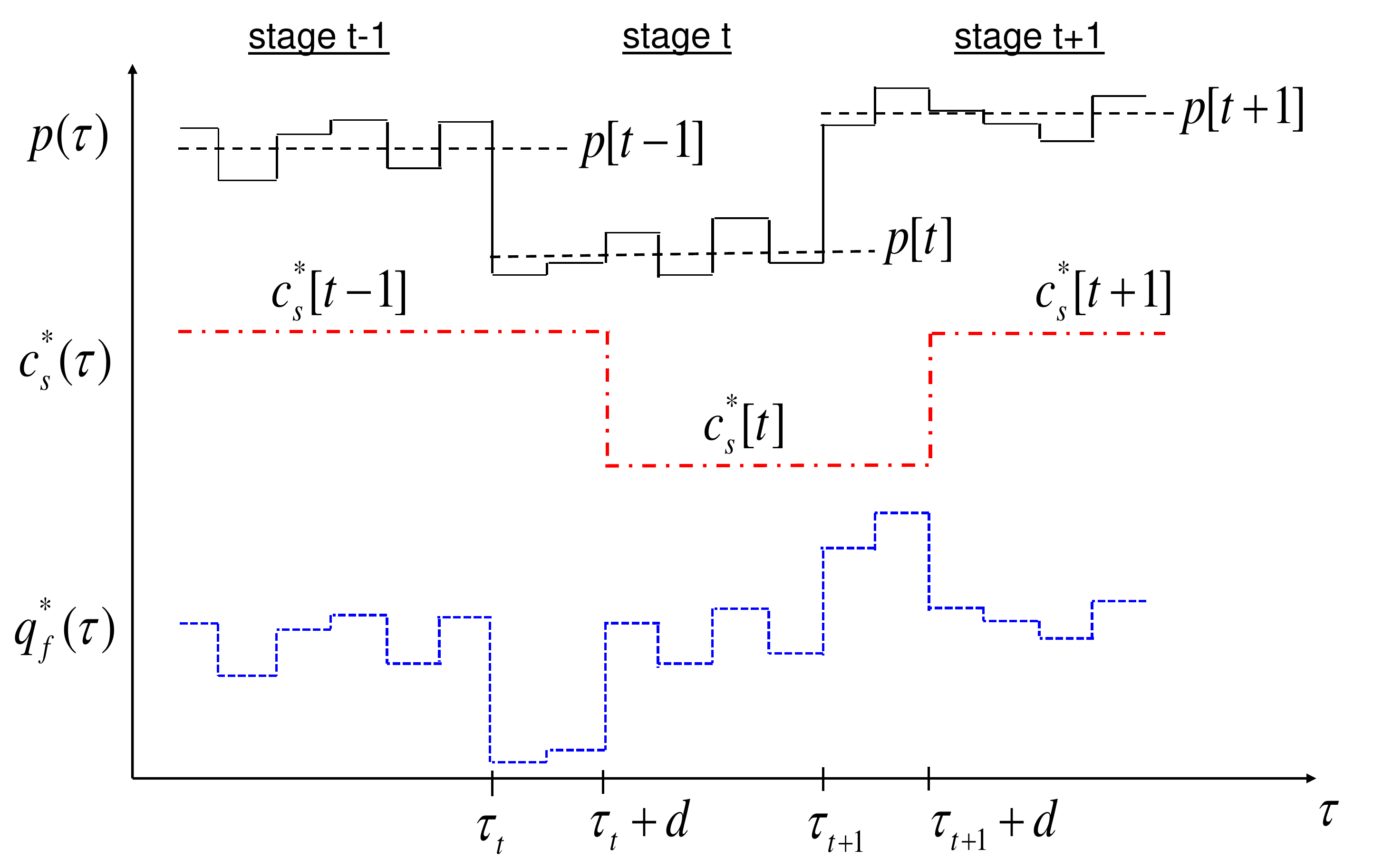}
\caption{Time line of voltage control, which is broken into stages with significantly different average load power $p[t]$. The transition between stages $t-1$ and $t$ occurs at time $\tau_t$. Following this transition, a new optimal output $c^*_s[t]$ of the switchable capacitor is computed but not implemented until a time delay $d$ after the transition. The D-STATCOM output $q_f^*(\tau)$ is computed and implemented at every time when $p(\tau)$ changes, which is especially important for voltage regulation during the interval from $\tau_t$ to $\tau_t+d$.}
\label{fig:timeline}
\end{figure}

The control is performed at two timescales: at slow timescale the capacitor $c_s$ is switched at most once per stage, and at fast timescale the D-STATCOM $q_f$ may be adjusted at every time $\tau$ when a new sample $p(\tau)$ is measured.
We assume there is a fixed time delay $d \in \mathbb{N}$ in capacitor switching, and $d < T_t$ for all $t$. This delay complicates the control time line, as demonstrated in Fig.~\ref{fig:timeline}.
During the bulk of stage $t-1$, the control output of the switchable capacitor is a constant $c_s^*[t-1]$. At the beginning time $\tau_t$ of the next stage $t$, a large change in $p$ occurs. A new control output $c_s^*[t]$ of the switchable capacitor is computed using the average load power $p[t]$ of stage $t$ (which is assumed to be known at $\tau_t$ if we consider the problem offline; see Section \ref{sec:heuristic_for_online_implementation} for the online case). However, due to the delay in switchable capacitor operation, $c_s(\tau)$ cannot change from $c_s^*[t-1]$ to $c_s^*[t]$ until $\tau_t+d$. A similar delay will occur after the transition from stage $t$ to $t+1$. For the D-STATCOM, however, a control output $q_f(\tau)$ is computed based on $p(\tau)$ and $c_s^*(\tau)$ and implemented instantaneously at every time $\tau$.

In the time line depicted in Fig.~\ref{fig:timeline}, the control of $c_s$ is coupled across every two consecutive stages. Specifically, the computation of $c_s^*[t]$ should incorporate a prediction about the behavior of $p[t+1]$ to limit the probability of unacceptable voltage deviations during the capacitor switching delay period $\tau_{t+1}$ to $\tau_{t+1}+d$, otherwise the uncertainty in $p[t+1]$ and the finite output range of $q_f$ could easily lead to a situation where the voltage at the load exceeds acceptable bounds in this period.
Given $p[t]$, the prediction of $p[t+1]$ is based on the transition probability $\pi\left(\cdot\left|\right.p[t]\right)$ introduced in Section \ref{subsec:loadmodel}.

For simplicity, we write $p[t]$, $p[t+1]$, $c_s[t]$ as $p$, $p^+$, $c_s$. Then (the offline version of) the slow-timescale capacitor control problem, denoted $\controlproblem_s(p,C_0,C_s,\overline q_f)$, is
\bq
&\min_{c_s, q_f, q_f^+} &  [i\left(p,C_0,c_s,q_f\right)]^2 r \label{eq:control_obj}\\
& \mbox{  s.t.  }& -\epsilon \leq [v\left(p,C_0,c_s,q_f\right)]^2 -v_0^2 \leq \epsilon \label{eq:constr:voltage_hard}\\
&& \text{Pr}\left\{[v(p^+,C_0, c_s, q_f^+)]^2 - v_0^2 \geq \epsilon \left|\right. p\right\} \leq \delta  \label{eq:constr:chance_high}\\
& & \text{Pr}\left\{[v(p^+,C_0, c_s, q_f^+) ]^2- v_0^2 \leq -\epsilon \left|\right. p\right\} \leq \delta \label{eq:constr:chance_low}\\
&& c_s \in \{\frac{k}{K}C_s | k = 0,...,K \} \label{eq:constr:sizes_cs}  \\
&& - \overline q_f \leq q_f \leq \overline q_f, \quad
- \overline q_f \leq q_f^+ \leq \overline q_f \label{eq:constr:sizes_qf}
\eq
where $i$ and $v$ as functions of $(p,C_0, c_s, q_f)$ or $(p^+,C_0, c_s, q_f^+)$ are specified by the DistFlow equations \eqref{eq:DF_l}--\eqref{eq:DF_v}. With respect to the average load power of the current stage, the objective \eqref{eq:control_obj} minimizes real power loss, and the deterministic constraint \eqref{eq:constr:voltage_hard} regulates voltage.
Chance constraints \eqref{eq:constr:chance_high}\eqref{eq:constr:chance_low} limit the probability of voltage violations during the capacitor switching delay period of the next stage, by incorporating transition probability from $p$ to $p^+$. Instead of limiting the voltage magnitude, we choose to limit its square, which simplifies the analysis below. Constraint \eqref{eq:constr:sizes_cs} specifies a discrete feasible set of $c_s$. Note that variables $q_f, q_f^+$ are not actual control actions of the D-STATCOM, but rather used to guarantee the existence of feasible (satisfying \eqref{eq:constr:sizes_qf}) operation points of the D-STATCOM in the current stage and in the capacitor switching delay period of the next stage, when we optimize over $c_s$.

At time $\tau_t$, problem $\controlproblem_s(p[t],C_0,C_s,\overline q_f)$ is solved for $c_s^*[t]$. The actions of the switchable capacitor are $c_s^*(\tau) = c_s^*[t]$ for $\tau_{t}+d \leq \tau <\tau_{t+1} + d$. Then at every time $\tau$, with $c_s^*(\tau)$ known, a fast-timescale D-STATCOM control problem $\controlproblem_{f}(p(\tau),C_0,c_s^*(\tau),\overline q_f)$ is solved for $q_f^*(\tau)$. For simplicity, write $p(\tau)$, $c_s^*(\tau)$, $q_f(\tau)$ as $p$, $c_s^*$,  $q_f$. Then $\controlproblem_{f}(p,C_0,c_s^*,\overline q_f)$ is
\bq
&\min_{q_f} & \quad  [i\left(p,C_0,c_s^*,q_f\right)]^2 r \label{eq:control_obj_cf}\\
& \mbox{  s.t.  }~& -\epsilon \leq [v\left(p,C_0,c_s^*,q_f\right)]^2 -v_0^2 \leq \epsilon \label{eq:constr:voltage_hard_cf}\\
&& - \overline q_f \leq q_f \leq \overline q_f \label{eq:constr:sizes_qf_cf}
\eq
where $q_f$ is optimized to minimize power loss while regulating the voltage at the current instant. While $\controlproblem_{s}$ is a chance-constrained OPF problem, $\controlproblem_{f}$ is a simpler OPF problem without chance constraints.

For both the capacitor and the D-STATCOM control problems above, the objective is to minimize power loss as long as voltage violations are avoided, which is common in practice and also makes sense for this specific system with a large HPC facility (potentially large associated power loss) and a simple circuit. Indeed, different control objectives might be chosen for different systems. For example, in a more complex distribution network with multiple loads, the objective might be finding a particular voltage profile across the network to minimize the total energy consumption, through mechanisms like Conservative Voltage Reduction \cite{farivar2011inverter, farivar2012optimal, schneider2010evaluation}.

\subsection{Optimal sizing of reactive power sources}

The optimal objective value of $\controlproblem_s(p,C_0, C_s, \overline{q}_f)$, i.e., the minimum power loss with respect to the average load power $p$ of a stage, is denoted by $L(p,C_0,C_s,\overline{q}_f)$. When planning the sizes of reactive power devices that will be installed in the circuit, we need account for the cost of the expected minimum power loss and the capital cost of devices. Hence, an optimal sizing problem is formulated as
\bq
\min_{\left(C_0,  C_s,  \overline q_f\right) \in \feasiblesize}&\quad& k_p \int_{\underline p}^{\overline p} L(p,C_0,C_s,\overline{q}_f) \rho(p) dp  \nonumber
\\
&&+ L_0(C_0)+L_s(C_s) + L_f(\overline q_f).
\label{obj:sizing}
\eq
The integral term in \eqref{obj:sizing} is the expectation of minimum power loss resulting from the capacitor control $\controlproblem_s$. Note that though $\controlproblem_s$ takes average load powers of stages as input, the integral in \eqref{obj:sizing} is taken over the stationary distribution $\rho(\cdot)$ of load powers sampled at 5-second timescale, since the stationary distribution of average stage powers does not include information of durations of stages.

In \eqref{obj:sizing}, the coefficient $k_p$ converts the expected power loss into a cost which has the same unit as the capital costs $L_0$, $L_s$ and $L_f$ of the fixed capacitor, the switchable capacitor and the D-STATCOM.  Let $\mathbb{R}_0^+$ denote the set of non-negative real numbers. The domain $\feasiblesize$ of the optimal sizing problem is the set of points $(C_0, C_s,  \overline q_f) \in (\mathbb{R}_0^+)^3$ such that $\controlproblem_s(p,C_0,C_s,\overline q_f)$ is feasible for all $p \in  [\underline p,~\overline p]$.

\section{Heuristic solution and implementation}\label{sec:allheuristics}

\subsection{Difficulties and sketch of approach}\label{susec:ideas_difficulties}

We formulated our control problem as two OPF problems on a one-branch, single-phase circuit, which are usually simple to solve. However, extra difficulties are brought in by the optimal sizing problem, since neither its objective function \eqref{obj:sizing} nor its domain $\feasiblesize$ has a closed-form expression, which makes it hard to solve with analytical methods. Probabilistic metaheuristics, e.g., simulated annealing, genetic algorithm and particle swarm optimization, are considered good candidate numerical methods to search for a (usually approximate) globally optimal solution for the sizing problem. In the rest of this paper we use simulated annealing (SA) \cite{kirkpatrick1983optimization}, but the techniques we develop can be applied to other metaheuristics in a same or similar way.

A key process in SA is to evaluate the objective value and, in particular, the integral term in \eqref{obj:sizing} for any given $(C_0, C_s, \overline q_f)$. In practice, we use the numerical approximation
\bq\label{eq:integral_approx}
\sum\limits_{n =1}^{N} L(p_n,C_0,C_s,\overline{q}_f) \tilde\rho_n
\eq
where $\underline p = p_0 < p_1<...<p_N = \overline p$ is a partition of the interval $\left[\underline p ,~ \overline p\right]$, and $\tilde\rho_n = \int_{p_{n-1}}^{p_n} \rho(p) d p$ is the probability that the real power load lies in the subinterval $\left[p_{n-1},~p_n\right]$. If $\controlproblem_s(p_n,C_0,  C_s,  \overline q_f)$ is infeasible for any $n \in \{1,...,N\}$, then $(C_0,  C_s, \overline q_f) \notin \feasiblesize$ and is assigned an infinitely high objective value for the sizing problem. If $\left(C_0,  C_s, \overline q_f\right) \in \feasiblesize$, evaluation of the integral term in \eqref{obj:sizing} requires solving $\controlproblem_s$ for $N$ times, one for each bin in the approximation in \eqref{eq:integral_approx}.   If $N$ is large, the computation of the objective value becomes expensive. Moreover, the chance constraints \eqref{eq:constr:chance_high}\eqref{eq:constr:chance_low} do not have closed-form expressions, making it more complex to solve $\controlproblem_s(p_n,C_0,  C_s,  \overline q_f)$.

To reduce the computational burden in our SA-based approach, we make simplifications for the underlying capacitor control problems and develop a heuristic to approximately solve them by exploiting the structure of the simplified problems. By doing this the evaluation of the integral term in \eqref{obj:sizing} is simplified, and we develop a computationally efficient heuristic to solve the optimal sizing problem. We also design a heuristic to implement the voltage control proposed in Section \ref{subsec:optimal_var_control} in an online manner in real time. Below we describe the details of our approach.

\subsection{Heuristic for capacitor control}

Indeed, from \eqref{eq:DF_l}--\eqref{eq:DF_v} one can solve for $v$ and $i$ explicitly in closed forms of $(p,C_0, c_s, q_f)$, using the classic formula of roots of quadratic equations. However, these explicit solutions still take such complicated forms that solving for $\controlproblem_s$ is computationally expensive. Hence we perform the following approximations to obtain a simplified version of $\controlproblem_s$, which has a clearer structure of how the solution depends on the input.

First, we simplify the expression of $i^2$. In a realistic distribution circuit, the real and reactive power losses $i^2 r $ and $i^2 x$ are much smaller than the sending-end real and reactive powers $P$ and $Q$, respectively. Hence by \eqref{eq:DF_l}--\eqref{eq:DF_Q} we have
\bq
i^2 \approx \frac{p^2 + \left( v^2 f_0  (C_0+ c_s) +  q_f - \phi p\right)^2}{v_0^2}. \label{eq:approx_l}
\eq

Second, we convert constraint \eqref{eq:constr:voltage_hard} into affine inequalities in $(c_s, q_f, p,i^2 )$. From \eqref{eq:DF_P}--\eqref{eq:DF_v} we have
\bq\label{eq:explicit_v}
v^2 = \frac{v_0^2 - 2(r+\phi x) p + 2x q_f - i^2 (r^2 + x^2) }{1-2x f_0 \left(C_0 + c_s\right)}.
\eq
We only consider $1-2xf_0 \left(C_0 + c_s\right) >0$ since it is a stability requirement that an increase in reactive power injection (or equivalently, in $C_0$, $c_s$ or $q_f$) results in an increase in voltage magnitude \cite{andersson2004modelling}. Substituting \eqref{eq:explicit_v} into \eqref{eq:constr:voltage_hard}, we have
\bq
&&q_{\text{vc},1}\left(c_s,q_f\right):=f_0 \left(v_0^2 + \epsilon \right) (C_0 + c_s) + q_f   \nonumber\\
&\leq& \left(\frac{r}{x}+ \phi \right) p + \frac{\epsilon}{2x} + \frac{i^2 \left(r^2 + x^2 \right)  }{2x}=: g_1 (p,i^2),
\label{eq:hard_constr_reactivepower_1}
\\
&&q_{\text{vc},2}\left(c_s,q_f\right):=f_0 \left(v_0^2 - \epsilon \right) (C_0 + c_s)  + q_f \nonumber \\
&\geq& \left(\frac{r}{x}+ \phi \right) p - \frac{\epsilon}{2x} + \frac{i^2 \left(r^2 + x^2 \right)  }{2x}=: g_2 (p,i^2) \label{eq:hard_constr_reactivepower_2}
\eq
where the left-hand-sides and right-hand-sides are affine functions of $(c_s, q_f)$ and affine functions of $(p,i^2)$.

Third, we convert chance constraints \eqref{eq:constr:chance_high}\eqref{eq:constr:chance_low} into simpler deterministic constraints. Similar to how we obtained \eqref{eq:hard_constr_reactivepower_1}\eqref{eq:hard_constr_reactivepower_2} from \eqref{eq:constr:voltage_hard}, we have, from \eqref{eq:constr:chance_high}\eqref{eq:constr:chance_low}, that
\bq\label{eq:chance_constr_reactivepower_1}
\text{Pr}\left( q_{\text{vc},1} (c_s,q_f^+)  \geq g_1(p^+,(i^+)^2) \left|\right. p\right) &\leq& \delta \\
\label{eq:chance_constr_reactivepower_2}
\text{Pr}\left(q_{\text{vc},2} (c_s,q_f^+)  \leq g_2(p^+,(i^+)^2 ) \left|\right. p\right) &\leq& \delta
\eq
where $i^+$ denotes $i(p^+,C_0,c_s,q_f^+)$. Given the current-stage average load power $p$, find two powers $\tilde{h}_1$ and $\tilde{h}_2$ as
\bq
\nonumber
\tilde h_1 (p) &:=& \sup\{\left.h\in[\underline p,~\overline p] \right|  \int_{\underline p}^{h} \pi \left(p^+|p\right) dp^+ \leq \delta\} \\
\nonumber
\tilde h_2 (p) &:=&\inf\{\left.h\in\left[\underline p,~\overline p\right] \right|  \int_{h}^{\overline p} \pi \left(p^+|p\right) dp^+ \leq \delta\}.
\eq
Then from \eqref{eq:chance_constr_reactivepower_1}\eqref{eq:chance_constr_reactivepower_2}, the chance constraints are converted into deterministic constraints:
\bq
q_{\text{vc},1} (c_s,q_f^+)
&\leq&  \left( \frac{r}{x} + \phi\right) \tilde h_1(p) + \frac{\epsilon}{2x}+ \frac{r^2 + x^2}{2x}  (i^+)^2 \nonumber
\\
&=:& h_1(p, (i^+)^2) \label{neq:chance_constr_1}
\\
q_{\text{vc},2} (c_s,q_f^+)  &\geq&  \left( \frac{r}{x} + \phi \right) \tilde h_2(p) - \frac{\epsilon}{2x}+ \frac{r^2 + x^2}{2x} (i^+)^2\nonumber
\\
&=:&  h_2(p,(i^+)^2). \label{neq:chance_constr_2}
\eq

Fourth, we approximate the $\left(p,C_0,c_s,q_f\right)$-dependent argument $i^2$ in $g_1(p,i^2)$ and $g_2(p,i^2)$ in \eqref{eq:hard_constr_reactivepower_1}\eqref{eq:hard_constr_reactivepower_2} with a constant $\tilde l$.\footnote{We abuse the notation by letting $\tilde l$ denote a vector in Section \ref{subsec:heuristic_sizing}. Its meaning should be clear given the context.}
Moreover we replace the $(p^+,C_0,c_s,q_f^+)$-dependent argument $(i^+)^2$ in $h_1(p, (i^+)^2)$ in \eqref{neq:chance_constr_1} with its estimated lower bound $\underline l^+$, and replace $(i^+)^2$ in $h_2(p, (i^+)^2)$ in \eqref{neq:chance_constr_2} with its estimated upper bound $\overline l^+$, where both $\underline l^+$ and $\overline l^+$ are constant.
Section \ref{subsec:heuristic_sizing} explains the way we obtain $\tilde l$, while $\underline l^+$ and $\overline l^+$ are estimated as follows. Suppose \eqref{eq:constr:voltage_hard}\eqref{eq:hard_constr_reactivepower_1}\eqref{eq:hard_constr_reactivepower_2} are also satisfied when $(p, q_f)$ is replaced by $(p^+, q_f^+)$ (which indeed occurs with a high probability $1-2\delta$). Then \eqref{eq:approx_l}, which also holds when $(i, p, q_f)$ is replaced by $(i^+, p^+, q_f^+)$, implies
\bq
(i^+)^2 &\geq& \frac{1}{v_0^2}\left[\underline p ^2 + \left(\frac{r}{x}\cdot \underline p - \frac{\epsilon}{2x} + \frac{\left(r^2 + x^2 \right)}{2x} (i ^+)^2\right)^2\right] \nonumber
\\
&\approx&  \frac{1}{v_0^2}\left[\underline p ^2 + \left(\frac{r}{x}\cdot \underline p - \frac{\epsilon}{2x} \right)^2\right]  =: \underline l^+, \nonumber
\\
(i^+)^2 &\leq&\frac{1}{v_0^2} \left[\overline p^2 + \left(\frac{r}{x}\cdot \overline p + \frac{\epsilon}{2x} + \frac{\left(r^2 + x^2 \right)}{2x} (i ^+)^2 \right)^2\right] \nonumber
\\
&\approx&\frac{1}{v_0^2} \left[\overline p ^2 + \left(\frac{r}{x}\cdot \overline p + \frac{\epsilon}{2x} \right)^2\right]  =: \overline l^+ \nonumber
\eq
where the approximate equalities result from dropping the term associated with relatively small power loss.
A significant component of these simplifications is that $g_1(p, \tilde l)$, $g_2(p,\tilde l)$, $h_1(p,\underline l^+)$, $h_2(p, \overline l^+)$ are known a priori when $p$ is given. Hence, with those terms on the right-hand-sides, inequalities \eqref{eq:hard_constr_reactivepower_1}\eqref{eq:hard_constr_reactivepower_2}\eqref{neq:chance_constr_1}\eqref{neq:chance_constr_2} become simple affine constraints in $(c_s, q_f, q_f ^+)$.

The four steps of approximations above render us a simple way to approximately solve $\controlproblem_s$, which is to minimize \eqref{eq:approx_l} subject to \eqref{eq:hard_constr_reactivepower_1}\eqref{eq:hard_constr_reactivepower_2} (with $i^2$ replaced by a constant $\tilde l$), and \eqref{neq:chance_constr_1}\eqref{neq:chance_constr_2} (with $(i^+)^2$ replaced by constants $\underline l^+$ and $\overline l^+$, respectively), and \eqref{eq:constr:sizes_cs}\eqref{eq:constr:sizes_qf}. A further observation is that the objective \eqref{eq:approx_l} of the simplified problem is decreased by decreasing $c_s$ and $q_f$. Indeed, in practice $\epsilon$ is selected to be much smaller than $2r\underline p$, which makes $v^2 f_0  (C_0 + c_s) + q_f  > \phi p$ by \eqref{eq:constr:voltage_hard}\eqref{eq:hard_constr_reactivepower_2}. Moreover, decreasing $c_s$ and $q_f$ results in an decrease in $v^2$ by \eqref{eq:explicit_v}.

Hence we design the following heuristic $\heuristiccontrol$ to approximately solve the capacitor control problem $\controlproblem_s(p,C_0, C_s,  \overline q_f)$ and get $\tilde L(p, C_0, C_s, \overline q_f;\tilde l)$, an approximation of the actual optimal objective $L(p, C_0, C_s, \overline q_f)$. Moreover, a variable ``$\text{feasibility\_flag}$'' is set to be $1$, which means $\controlproblem_s(p,C_0, C_s,  \overline q_f)$ is feasible, if a $c_{s}^*$ is found by the heuristic, and $0$ otherwise.

\begin{heu}{\emph{$\heuristiccontrol(p, C_0, C_s, \overline q_f;\tilde l)$: capacitor control}}\\
\noindent $\text{feasibility\_flag} = 0$;\\
\noindent \textbf{for} $k = 0,1,...,K $ \textbf{do}\\
\indent\textbf{if} $q_{\text{vc},1} \left(\frac{k}{K}C_s,- \overline q_f\right) \leq \min\left (g_1(p,\tilde l),h_1(p, \underline l^+)\right)$ \\
\indent and $q_{\text{vc},2}\left(\frac{k}{K}C_s, \overline q_f\right)   \geq \max\left (g_2(p,\tilde l),h_2(p, \overline l^+)\right)$ \textbf{do}\\
\indent\quad$\text{feasibility\_flag} = 1$;\\
\indent\quad$c_{s}^* = \frac{k}{K}C_s$;\\
\indent\quad$q_{f}^* = \max\left(-\overline q_f,g_2(p,\tilde l) - f_0 \left(v_0 - \epsilon\right) \left(C_0 + c_{s}^* \right) \right)$;\\
\indent\quad$\tilde L(p, C_0, C_s, \overline q_f;\tilde l) =  [i(p,C_0,c_{s}^*,q_{f}^*)]^2 r $;\\
\indent\quad \textbf{return};\footnote{In pseudo codes of this paper ``return'' means terminating the current heuristic and returning the values of all variables computed.}\\
\indent\textbf{end if};\\
\noindent \textbf{end for};\\
\noindent \textbf{return};
\end{heu}

Note that the result of $\heuristiccontrol$ depends on the constant $\tilde l$, whose selection will be explained below. The heuristic $\heuristiccontrol$ forms a basis for developing the heuristic to solve the optimal sizing problem.

\subsection{Heuristic for optimal sizing}\label{subsec:heuristic_sizing}

Suppose a partition $\underline p = p_0 < p_1<...<p_N = \overline p$ is given and fixed, and for every $n \in \{1,...,N\}$ the probability $\tilde\rho_n$ of load power lying in the subinterval $\left[p_{n-1},~p_n\right]$ is known. When solving the optimal sizing problem with SA, the function $\tilde E (C_0, C_s, \overline q_f;\tilde l)$ below is used to approximate the objective value \eqref{obj:sizing} at a given point $(C_0, C_s, \overline q_f)$, where $\tilde l$ is a vector $(\tilde l_1,...,\tilde l_N)$ of constants used to approximate the minimum value of $i^2$ for each input $p_n$ to the underlying problem $\controlproblem_s$. Note that $\tilde E (C_0, C_s, \overline q_f;\tilde l)$ is assigned an extremely high value as $+\infty$, i.e., $ (C_0, C_s, \overline q_f)$ is marked as infeasible, if $\heuristiccontrol(p_n, C_0, C_s, \overline q_f;\tilde l_n)$ for any $n \in  \{1,...,N\}$ returns $\text{feasibility\_flag}_n =0$.

\begin{heu}{\emph{$\tilde E(C_0, C_s, \overline q_f;\tilde l)$: approximate sizing objective}}
\\
\noindent\textbf{for} $n = 1,...,N$ \textbf{do}
\\
\indent Run $\heuristiccontrol(p_n, C_0, C_s, \overline q_f;\tilde l_n)$;
\\
\indent\textbf{if} $\text{feasibility\_flag}_n == 0$ \textbf{do}
\\
\indent\indent$\tilde E(C_0, C_s, \overline q_f;\tilde l) = +\infty$;
\\
\indent\indent\textbf{return};
\\
\indent\textbf{end if};
\\
\noindent \textbf{end for};
\\
\noindent $\tilde E(C_0, C_s, \overline q_f;\tilde l)  = k_p \sum\limits_{n =1}^{N} \tilde L(p_n, C_0, C_s, \overline q_f;\tilde l_n)  \tilde\rho_n    + L_0(C_0)+L_s(C_s) + L_f(\overline q_f)$;
\\
\noindent \textbf{return};
\end{heu}
Based on the approximate objective function above, an iterative heuristic $\heuristicsizing$ is developed to approximately solve the optimal sizing problem. In the $j$-th iteration, $\heuristicsizing$ runs SA with objective function $\tilde E(\cdot;\tilde l^{*,j})$ to obtain an optimal $(C_0^{*,j},C_s^{*,j}, \overline q_f^{*,j})$. Based on the outputs of underlying $\heuristiccontrol$ heuristics, $\tilde l^{*,j}$ is updated to $\tilde l^{*,j+1}$.
\begin{heu}{\emph{$\heuristicsizing $: optimal sizing}}
\\
\noindent $j = 0$; $\tilde l^{*,0} =0$;
\\
\noindent \textbf{while} termination condition == false \textbf{do}
\\
\indent Run SA with $\tilde E(\cdot;\tilde l^{*,j})$ and get $(C_0^{*,j},C_s^{*,j}, \overline q_f^{*,j})$;
\\
\indent\textbf{for} $n=1,...,N$ \textbf{do}
\\
\indent\indent$\tilde l^{*, j+1}_n = \frac{1}{r}\cdot \tilde L(p_n, C_0^{*,j}, C_s^{*,j}, \overline q_f^{*,j};\tilde l^{*,j}_n)$;
\\
\indent\textbf{end for};
\\
\indent$j= j+1$;
\\
\noindent \textbf{end while};
\\
\noindent $(C_0^*,C_s^*, \overline q_f^*) = (C_0^{*,j},C_s^{*,j}, \overline q_f^{*,j})$;
\\
\noindent \textbf{return};
\end{heu}
\noindent An example of the termination condition is that some norms $\|\tilde l^{*,j+1}-\tilde l^{*,j}\|$, $\|C_0^{*,j+1}-C_0^{*,j}\|$, etc. are smaller than certain thresholds. In the numerical experiments in Section \ref{sec:results} this condition is always satisfied within a small number of iterations. The fact that only a small number of iterations are required and each iteration works on SA with a simple objective function indicates that $\heuristicsizing$ is computationally efficient in solving the optimal sizing problem.

\subsection{Heuristic for real-time control}\label{sec:heuristic_for_online_implementation}

In Section \ref{subsec:heuristic_sizing} we solved the optimal sizing problem. Now we suppose reactive power sources of optimal sizes $\left(C_0^*,C_s^*, \overline q_f^*\right)$ have been installed in the circuit and look at the implementation of real-time control. Recall that we formulated the capacitor control problem in Section \ref{subsec:optimal_var_control} in an offline manner, i.e., by assuming that the average load power $p[t]$ of stage $t$ is known at the beginning $\tau_t$ of stage $t$. This assumption, however, does not hold in practice since $p[t]$ also depends on inputs $p(\tau)$ for $\tau > \tau_t$. Therefore the heuristic for real-time control should be implemented online for sequential arrivals of input $\{p(\tau), \tau \in \mathbb{N}_0\}$.

To this end, we develop a heuristic $\heuristiconline$ which determines the starts of new stages online, estimates the average load powers of stages, solves $\controlproblem_s$ for capacitor control at every stage and solves $\controlproblem_f$ for D-STATCOM control at every time when the load power is sampled. Specifically, a threshold $p_{\text{th}}$ is used to determine the starting time $\tau_t$ of a stage $t$. At $\tau_t$ the controller takes $p(\tau_t)$ as an estimate of $p[t]$ and solves $\controlproblem_s(p(\tau_t), C_0^*, C_s^*, \overline q_f^*)$ for $c_s^*(\tau_t+d)$, due to the operation delay $d$ of the switchable capacitor. The estimate of $p[t]$ is updated for $\tau = \tau_t+1,...,\tau_t+T_t-1$. Whenever the updated estimate of $p[t]$ deviates from the previous input to $\controlproblem_s$ by more than a preset threshold $p_{\text{est}}$, problem $\controlproblem_s$ needs to be solved again with the updated estimate of $p[t]$ as new input. The D-STATCOM control problem $\controlproblem_f$ is a simple OPF problem and can be solved using standard techniques, which are beyond the scope of this paper.
Details of $\heuristiconline$ are given below. Suppose $\heuristiconline$ has been running for $\tau<0$ so that the values of $t$, $p[t]$, $\tilde p[t]$, $T_t$ and $c_s^*(\tau),...,c_s^*(\tau+d-1)$ are known at $\tau=0$.

\begin{heu}{\emph{$\heuristiconline$: real-time voltage control}}
\\
\noindent \textbf{for} $\tau = 0, 1,2,...$ \textbf{do}
\\
\indent \textbf{if} $\left|p(\tau) - p[t]\right| > p_{\text{th}}$ \textbf{do}
\\
\indent\indent $t = t+1$; $p[t] = p(\tau)$; $\tilde p[t] = p[t]$; $T_t =1$;
\\
\indent\indent Solve $\controlproblem_s\left(\tilde p[t], C_0^*, C_s^*, \overline q_f^*\right)$ for $c_s^*(\tau+d)$;
\\
\indent\textbf{else do}
\\
\indent\indent$p[t] = \frac{p[t]\times T_t +p(\tau)}{T_t+1} $; $T_t =T_t+1$;
\\
\indent\indent\textbf{if} $\left|p[t]-\tilde p[t]\right| > p_{\text{est}}$ \textbf{do}
\\
\indent\indent\indent$\tilde p[t] = p[t]$;
\\
\indent\indent\indent Solve $\controlproblem_s\left(\tilde p[t], C_0^*, C_s^*, \overline q_f^*\right)$ for $c_s^*(\tau+d)$;
\\
\indent\indent\textbf{else do}
\\
\indent\indent\indent$c_s^*(\tau+d) = c_s^*(\tau+d-1)$;
\\
\indent\indent\textbf{end if;}
\\
\indent\textbf{end if};
\\
\indent Solve $\controlproblem_f(p(\tau), C_0^*, c_s^*(\tau), \overline q_f^*)$;
\\
\noindent\textbf{end for};
\end{heu}

Note that the previous process of running $\heuristicsizing$ for optimal sizing has provided us a great deal of information to simplify computations in $\heuristiconline$. For example, for all $n \in \{1,...,N\}$ we have got $\tilde l_n^*$, $g_1(p_n,\tilde l_n^*)$, $g_2(p_n,\tilde l_n^*)$, $h_1(p_n,\underline l^+)$, $h_2(p_n,\overline l^+)$ and $c_{s,n}^*$. When solving $\controlproblem_s\left(\tilde p[t], C_0^*, C_s^*, \overline q_f^*\right)$ in real-time control $\heuristiconline$, if it happens that $\tilde p[t] = p_n$ for some $n$, then we know its solution is $c_{s,n}^*$ without actually solving it. Otherwise it can be solved by running $\heuristiccontrol(\tilde p[t], C_0^*, C_s^*, \overline q_f^*;\tilde l^*[t] )$, in which $\tilde l^*[t]$, $g_1(\tilde p[t],\tilde l^*[t])$, etc. can be obtained through interpolation of $\tilde l_n^*$, $g_1(p_n,\tilde l_n^*)$, etc. for $p_n$ neighboring $\tilde p[t]$. Such simplifications can accelerate the computations in real-time control. Indeed, in the numerical experiments in Section \ref{sec:results}, the computation time of running $\heuristiconline$ is negligible compared to the time step between consecutive control actions.

As an additional remark to this section, the multiple heuristics proposed above are inspired by the insight we obtain from the structure of the simplified problem resulting from a series of approximations to the original capacitor control problem. Rigorous analysis of the impact of those approximations and the performance of the proposed heuristics, e.g., sub-optimality bounds of $\heuristiccontrol$ and $\heuristicsizing$ and convergence rate of $\heuristicsizing$, is our future work.

\section{Numerical results}\label{sec:results}

We solve the optimal sizing problem with the proposed heuristic and run simulations to test the proposed real-time control. We also study the dependence of the optimal device sizes and the performance of the proposed control on the choice of parameter $\delta$, the tolerable probability of voltage violations due to transitions of load power.

We take the model in Fig. \ref{fig:HPC_config} with the following parameter values selected. The per unit base power is 1 kW, $v_0  = f_0 = 1~\text{pu}$, $\phi = 0.2$, and $r = x = 1.1 \times 10^{-5}~\text{pu}$. The parameter $\epsilon$ for voltage regulation is 0.02 pu, which allows the voltage magnitude $v$ to fluctuate between 0.99 pu and 1.01 pu.\footnote{We make the acceptable voltage range very tight to exercise the proposed schemes. Larger loads (or distributed generation) will cause larger voltage swings that are closer to realistic limits.} Suppose the capital costs of reactive power sources are 
\bq
&&L_0 (C_0) = k_0 v_0^2 f_0  C_0, \quad L_s (C_s) = k_s v_0^2 f_0  C_s, \nonumber
\\
&& L_f (\overline q_f) = k_f \overline q_f. \nonumber
\eq
The price of energy (that supplies the real power loss) is $\$50/\text{MWh}$. Both the prices of the fixed capacitor and the switchable capacitor, in terms of dollars spent on per unit reactive power injection under nominal voltage and frequency, are $\$1000/\text{Mvar}$. The price of D-STATCOM, in terms of dollars spent on per unit reactive power injection, is $\$100000/\text{Mvar}$. Suppose all the reactive power devices can be used for 30 years. The prices above are then converted to values of $k_p$, $k_0$, $k_s$ and $k_f$ such that the objective \eqref{obj:sizing} of the optimal sizing problem measures the cost in dollars every day. For the switchable capacitor $K=1$, i.e., it can switch to either $0$ or $C_s$.

From the four-day trace of load power in Fig.~\ref{fig:p_trace}, we use samples in the first three days as the training set to measure the transition probabilities $\pi$ between stages and the stationary distribution $\rho$ of load power samples.
We use different parameter $\delta$ in different cases of the experiments, where a case means the process of solving the optimal sizing problem using $\heuristicsizing$ and then, with the resulting optimal sizes of devices, implementing the real-time control $\heuristiconline$ on the load power trace in the last day from Fig.~\ref{fig:p_trace}.

\begin{figure*}[!t]
\centering
\subfigure[]
{\includegraphics[height=5.6 cm]{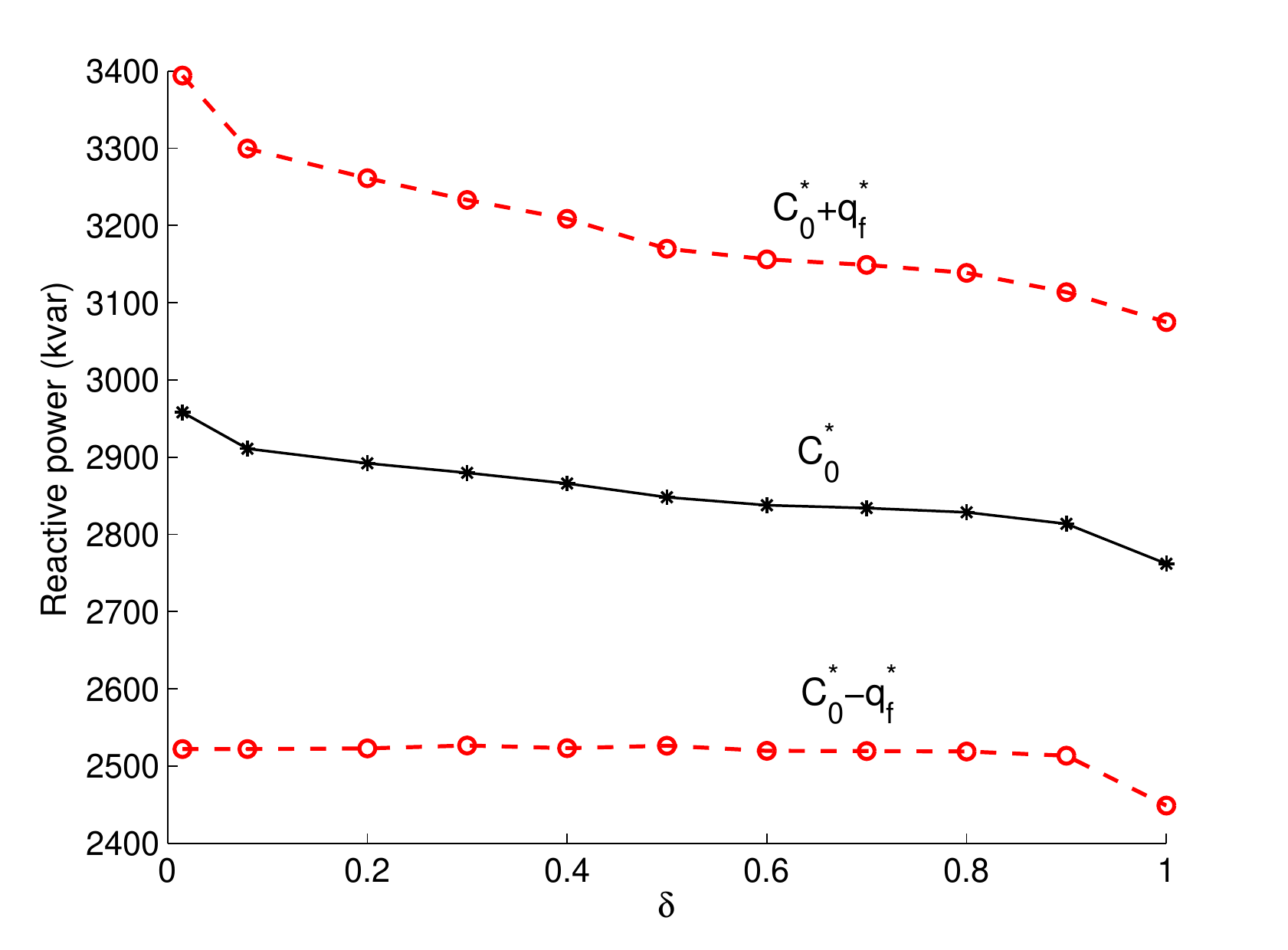}\label{fig:size_low}}
\hfil
\subfigure[]
{\includegraphics[height=5.6 cm]{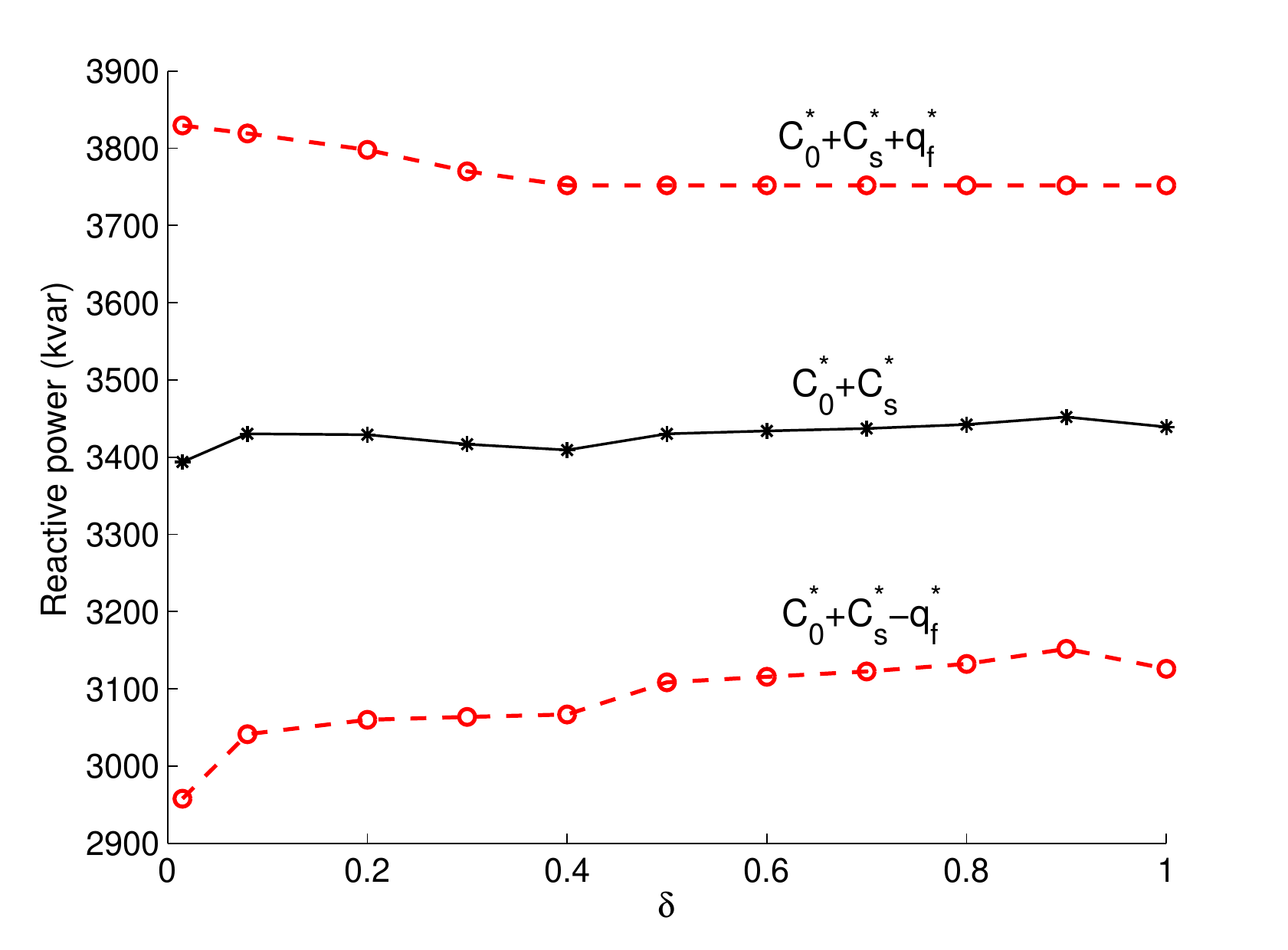}\label{fig:size_high}}
\caption{Sizes of control devices as functions of $\delta$. The range of aggregate reactive power injection of the fixed capacitor and the D-STATCOM is plotted for (a) $c_s =0$ and (b) $c_s = C_s^*$, respectively. }\label{fig:size}
\end{figure*}

Fig.~\ref{fig:size} shows the dependence of optimal device sizes on $\delta$.
Fig.~\ref{fig:size_low} shows $C_0^*$ and $C_0^*\pm \overline q_f^*$, i.e., the range of reactive power injection\footnote{The capacitance $C_0$ and its nominal reactive power injection $ v_0^2 f_0  C_0$ have the same per unit value, since $f_0 = v_0 =1~\text{pu}$.} when the switchable capacitor control $c_s=0$, which usually happens under low load power. On the other hand, Fig.~\ref{fig:size_high} shows $C_0^*+C_s^*$ and $C_0^*+C_s^* \pm \overline q_f^*$, i.e., the range of reactive power injection when $c_s=C_s^*$, which usually happens under high load power. We see in both subfigures that the range of reactive power injection gets broader as $\delta$ decreases, since with less tolerance of probabilistic voltage violations (smaller $\delta$), the D-STATCOM is required to have larger control capacity $\overline q_f^*$ to regulate voltage more safely when $c_s$ cannot switch immediately following a large transition of load power.
Another observation is that the lower bound $C_0^*-\overline q_f^*$ of the total control is almost constant for $\delta \leq 0.9$. Indeed, for those $\delta$, when $c_s =0$, the chance constraint \eqref{eq:constr:chance_high} is not binding in any capacitor control problem underlying the sizing problem.
For a similar reason $C_0^*+C_s^*+\overline q_f^*$ is almost constant for $\delta \geq 0.4$.

Note that we implement $\delta =1$ by removing the chance constraints \eqref{eq:constr:chance_high}\eqref{eq:constr:chance_low} in all the capacitor control problems underlying the sizing problem. Hence in Fig.~\ref{fig:size_low} there is a significant drop of the whole range of reactive power injection when $\delta$ is increased from $0.9$ to $1$. Indeed, after removing \eqref{eq:constr:chance_low} it is no longer necessary to maintain a high level of $C_0^* + q_f^*$ for voltage regulation during the capacitor switching delay period immediately after any possible large load increase, and thus the range of reactive power injection can be moved down to decrease power loss as well as capital cost. This decrease in power loss and capital cost, however, is obtained by suffering a higher risk of voltage violations, as shown in Figs.~\ref{fig:traj} and~\ref{fig:performance} below.

As sketched above we run the real-time control heuristic $\heuristiconline$ for many cases, each with a different $\delta$ and different device sizes depending on that $\delta$. For two of the cases with $\delta = 0.1$ and $\delta =1$, the real-time traces of voltage magnitude and real power loss are shown in Fig.~\ref{fig:traj}. The traces of voltage and power loss are also plotted for a benchmark case with only a fixed capacitor (whose size equals $C_0^*+C_s^*+\overline q_f^*$ when $\delta=0.1$) and no control.
\begin{figure*}[!t]
\centering
\subfigure[]
{\includegraphics[height=5.6 cm]{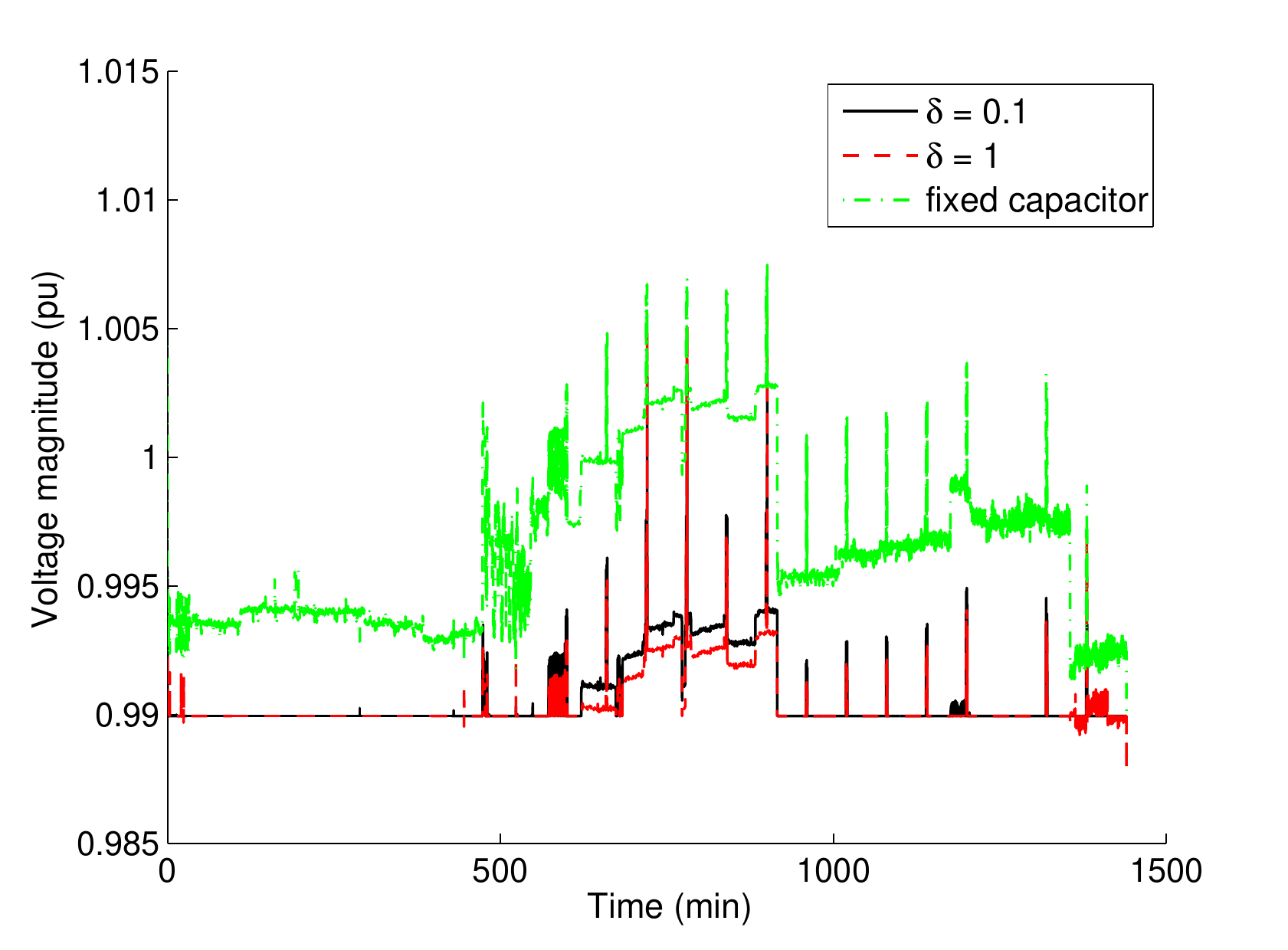}\label{fig:voltage_traj}}
\hfil
\subfigure[]
{\includegraphics[height=5.6 cm]{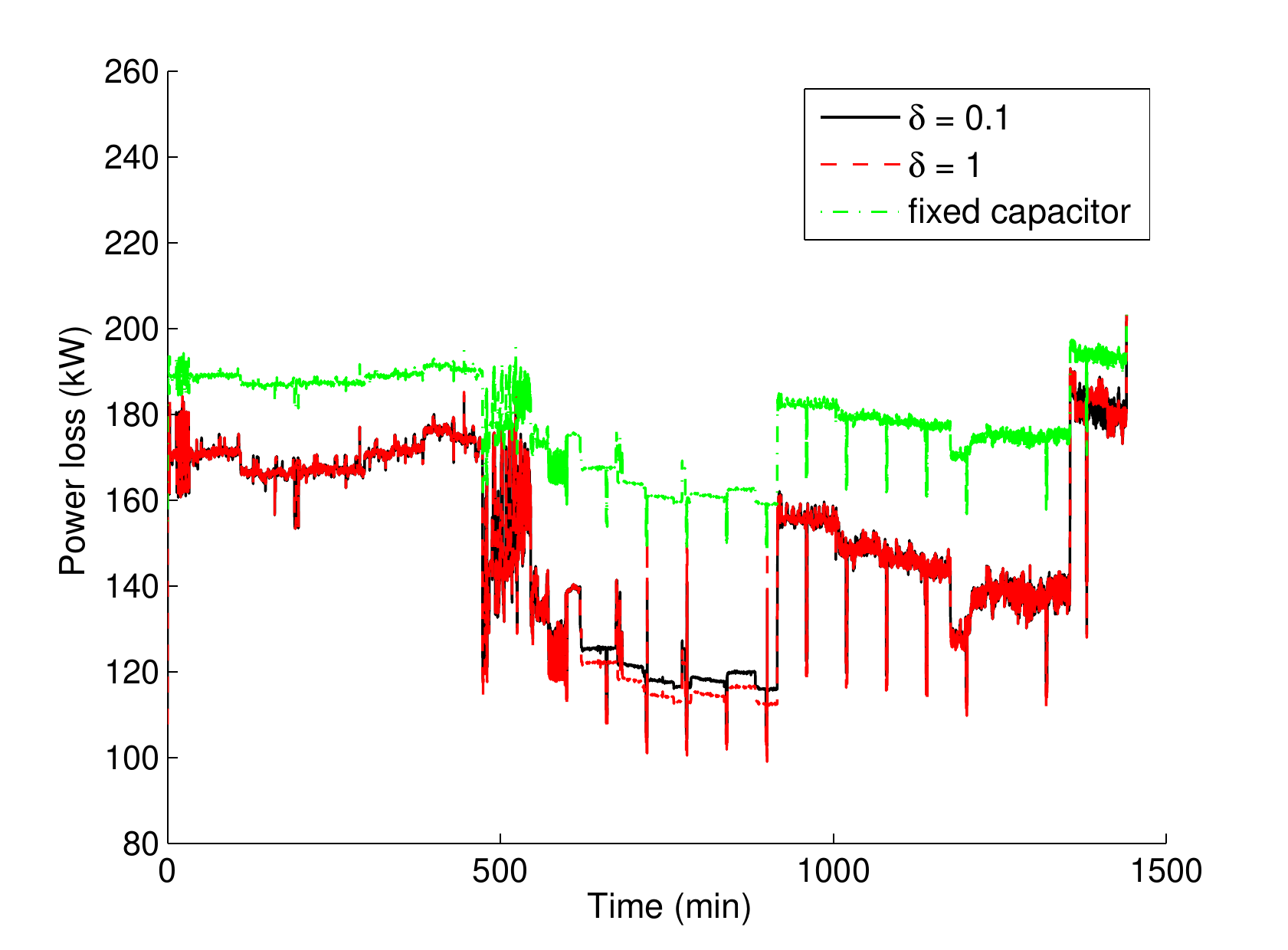}\label{fig:loss_traj}}
\caption{ Real-time traces of (a) voltage magnitude and (b) power loss for different $\delta$, and a benchmark case with only a fixed capacitor and no control.}\label{fig:traj}
\end{figure*}
We see that, in the benchmark case with only a fixed capacitor, the larger time-independent reactive power injection from the fixed capacitor results in higher voltages and losses at nearly all times compared to the two cases with controls. The case $\delta=0.1$ generally biases the voltage above the case $\delta =1$ (no chance constraints). This bias protects the system against experiencing an undervoltage when the load suddenly increases, as revealed near the end of the day when the voltage in the case $\delta = 1$ dips below 0.99 pu. This extra voltage safety provided by the chance constraints incurs increased power loss during periods when the case $\delta = 0.1$ biases the voltage up with additional reactive power injections.

For each case with different $\delta$, we record the proportion of 5-second samples in one day at which the voltage drops below 0.99 pu (indeed the voltage never swings above 1.01 pu so those recorded are all the samples with voltage violations). We also sum up the real-time power loss over one day, and add up the cost of that power loss and the average capital cost in one day.
Fig. \ref{fig:performance} shows the proportion of samples with voltage violations and the one-day total (capital plus power-loss-induced) cost for different $\delta$.
\begin{figure}[!t]
\centering
\includegraphics[height=5.6 cm]{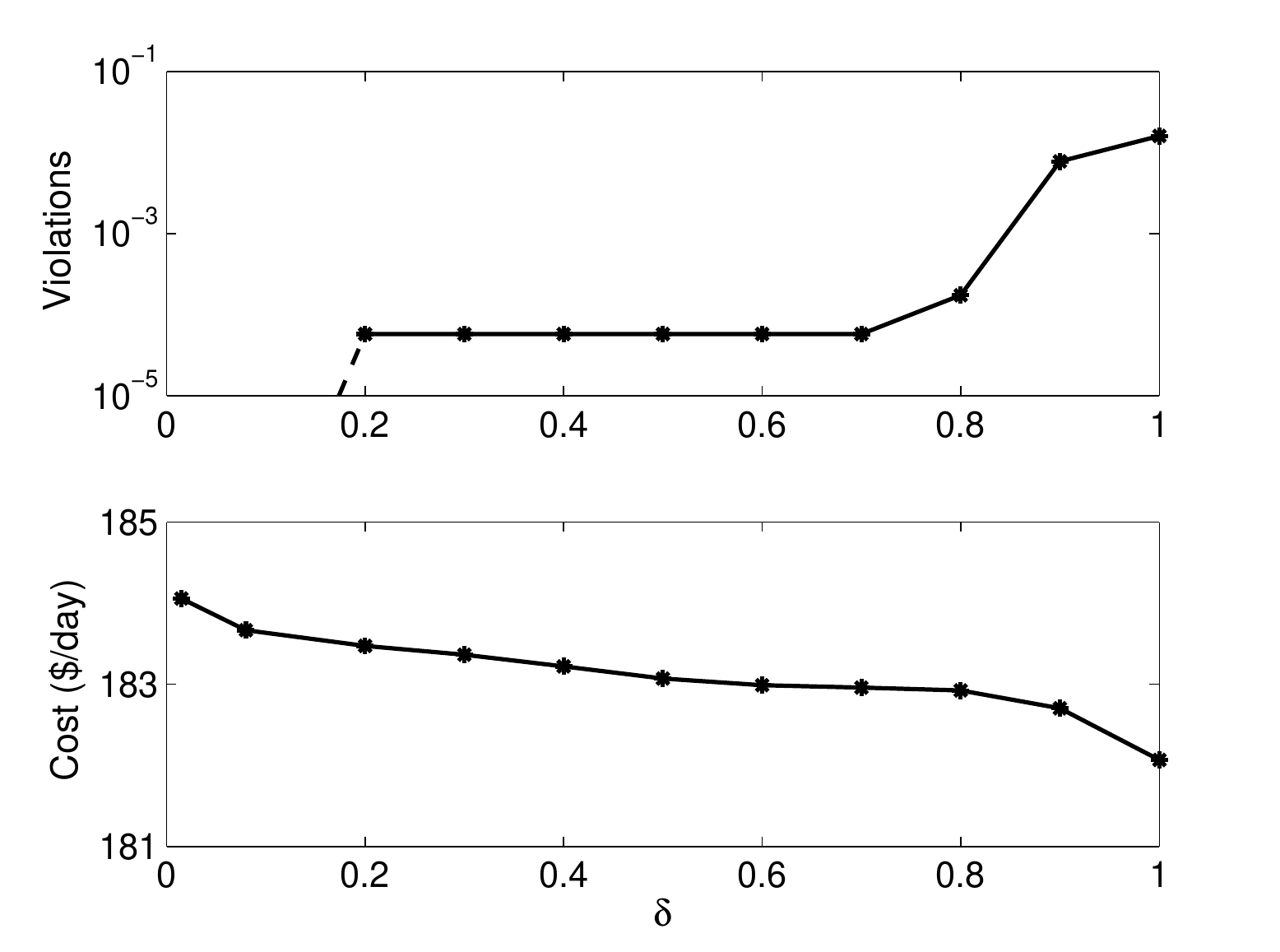}
\caption{Upper: the proportion of samples with voltage violations, which drops to zero when $\delta<0.2$. Lower: cost of the system in one day, including cost of power loss and capital cost. }
\label{fig:performance}
\end{figure}
As $\delta$ is decreased, the voltage control becomes more reliable as demonstrated by the significantly decreasing proportion of samples with voltage violations in the upper subfigure. The increased reliability only brings a modest increase in cost, as shown in the lower subfigure. Not shown in Fig.~\ref{fig:performance} is a benchmark case with only a fixed capacitor and no control. In that case the fixed capacitor is set high enough so that the voltage never drops below 0.99 pu, but the cost is as high as \$215/day due to the high power loss. We also consider another benchmark case in which there is only a D-STATCOM and there are no fixed and switchable capacitors. In this case the deterministic-constrained OPF problem $\controlproblem_{f}(p,C_0,c_s^*,\overline q_f)$ is solved every 5 seconds in real time with $C_0=c_s^*=0$ and $\overline q_f$ being the minimum value such that $\controlproblem_{f}$ is feasible for the peak load (and hence feasible all the time). The total cost is as high as \$207/day due to the high capital cost of the D-STATCOM. Therefore the cost for either benchmark case is much higher than the cost under the proposed control, whatever value $\delta$ is. 

As a main result of the experiments above, with the proposed heuristics to solve the optimal sizing problem and implement real-time control, the reliability of voltage regulation is significantly improved with moderate increase in cost, and hence a desired tradeoff can be achieved between performance of voltage regulation and cost efficiency.

\section{Conclusion}\label{sec:conclusions}
We have formulated a two-timescale optimization problem for joint control of a switchable capacitor and a D-STATCOM for voltage regulation in a distribution circuit with intermittent load. The slow-timescale capacitor control problem solves a chance-constrained OPF, which balances power loss with the probability of future voltage violations, by incorporating statistics of load changes over time. We have also integrated the result of the control problem into a sizing problem that determines the optimal sizes of reactive power sources. The optimal sizing problem allows a tradeoff between the expected cost due to power loss and the capital cost. We developed computationally efficient heuristics to solve the sizing problem and implement real-time control. In numerical experiments these heuristics were applied to measured data from an HPC load that routinely undergoes large and fast changes in power consumption. The results demonstrate the ability of the proposed schemes in  improving the reliability of voltage regulation with modest increase in cost.

This work is an initial step towards using chance constraints and load (or generation) statistics to size voltage control devices for distribution circuits. It can be extended to incorporate, e.g., multiple loads and multiple reactive power sources, tree-like circuits with multiple branches, multi-phase circuits, other loads or generation like PV generation, and multiple circuit configurations generated by distribution circuit switching.


\section*{Acknowledgment}

The authors would like to thank Steven Low for helpful discussions. This work was done during the internship of the first author at the Center for Nonlinear Studies, Los Alamos National Laboratory. This work was supported by the Advanced Grid Modeling Program within the Office of Electricity in the U.S. Department of Energy.

\IEEEtriggeratref{17}


%


\end{document}